\documentclass[jcp,preprint]{revtex4-1} 

\usepackage{amsmath}  
\usepackage{amsfonts} 
\usepackage{graphicx} 
\usepackage{epstopdf}
\begin{document}

\title{Confettti Ordering by Polymer Brushes}

\author{Galen T. Pickett}
\email{Galen.Pickett@csulb.edu} 
\affiliation{Department of Physics and Astronomy, California State University Long Beach, 1240 Bellflower Blvd., Long Beach, CA 90840, USA101} 


\date{\today}

\begin{abstract}
I consider the ordering of dilute platelet additives when incorporated
into an end-grafted polymer brush.
The competition between wetting interactions and the anisotropic stress
environment of the interior of the brush causes these platelet additives
to either remain suspended at the outer edge of the brush laying flat against 
the brush surface (as bits of confetti at rest on the ground), 
or to invade the interior of the brush in which case the
platelets stand end-on and in some cases protrude above the outer 
edge of the brush.
The orientation of the additives is controlled by the
ratio of the diameter of the additive to the thickness of the bare brush,
as well as the ratio of solvent-monomer and solvent-platelet interactions.
\end{abstract}

\maketitle 

%
%
%
%
%
%
\newcommand {\refto} {\cite}
\newcommand {\vol} {\bf}
\long\def\omit#1{}
\newcommand {\csi} {\xi}
\newcommand {\rhot} {\rho^{\prime}}
\newcommand {\csid} {\xi^{\prime}}
\newcommand {\gsim} {\stackrel{\textstyle >}{\sim}}
\newcommand {\lsim} {\stackrel{\textstyle <}{\sim}}
\newcommand {\vo} {v_o}
\newcommand {\vot} {v_o^{\prime}}
\newcommand {\aoverd} {a^{\prime}}
\newcommand {\peq} {d^3 P_{eq}}
%
%
%
%
%
%
%
\hyphenation{mon-o-mer mon-o-mers homo-poly-mer homo-poly-mers co-poly-mer
	co-poly-mers an-i-so-trop-ic macro-molecules den-dri-mers
	Den-dri-mer}

\section{Introduction}
\omit{
orinetation of block copolymers, control of microcopic pattern.
surface-active additives, surfactants, sam's of gold, orientation important,
tilt, geometry.
CNT long, asymetric.
inorganic exfoliated clay nanoparticles, orintation.
Brush, inherently anisotropic environment, use to control orientation
of additives. 
New properties, nanosurface composites.
outline of paper.
}
Self-assembly at the nanoscopic scale is responsible for an astonishing
variety of materials properties in functional surfaces \cite{self-assembly}.  
Homogeneously absorbed polymer layers are perhaps the simplest example,
where the trapped polymer layer dramatically improves the the ability 
of surfaces to avoid intimate contact (tribology)\cite{adsorbed} or selectively absorb
other target materials (cell-molecular lab-on-a-chip technology) \cite{cell_absorb}.
At the cost of designing more complex surface-active materials, 
more interesting functionality and patterns can be engineered to self-assemble.
Block copolymer thin films displaying an end-on ``perpendicular'' 
striped pattern, either through controlling competing wetting interactions \refto{pickett_perp}, 
through employing asymmetric cylinder-forming diblocks\refto{jaeger}, or through the
application of external electric fields are all examples of patterned
polymer films\refto{morkved} that can be used to order a third, absorbed, material.
Decorating these anisotropic patterns is a robust method for imparting a 
desired pattern to subsequent layers.

Central to ordering of the block copolymer layers is the subtle balance
between the stretching of the chains, creating an anisotropic stress
environment inside the brush, and substrate and solvent interactions between
monomers favoring the copolymer domains to conform to the substrate.  
Many avenues of controlling the orientation and texture of
the microsegregated patterns have been investigated \cite{bcp,bcp2}.

\begin{figure}
\centering
\includegraphics[scale=0.8]{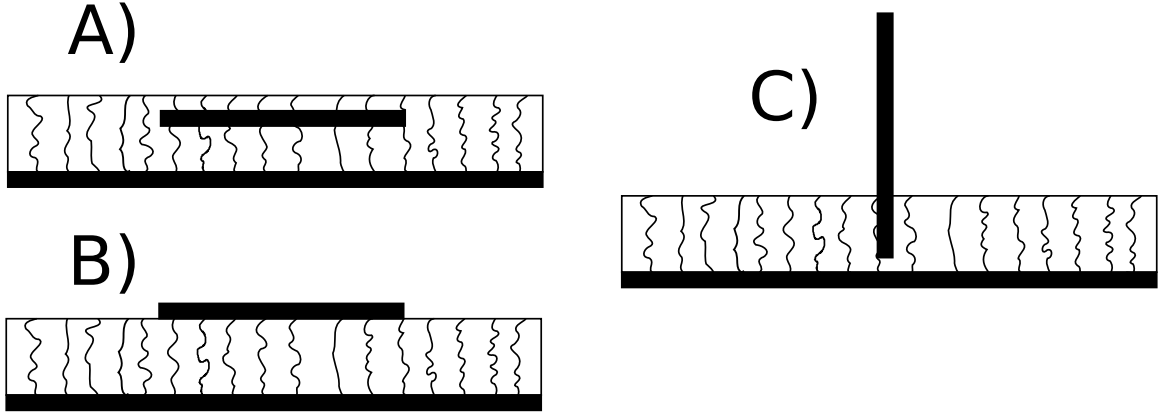}
\caption{
A) The platelet additive is immersed and wetted with brush polymers, the {\it wetted} confetti configuration.
B) The platelet additive rests on the surface of the brush, with an exposed dewetted surface, the {\it dewetted} confetti configuration.
C) The platelet additive rests perpendicular to the brush.
}
\label{schematic}
\end{figure}

Long, thin, additives, however, can act locally as substrates themselves,
so an interesting question is, what are the ordering boundary conditions on the hard surface of the additive?
A grafted polymer layer is composed of
anisotropic chains stretching away from the main substrate.
Long thin additives can interact with a brush by adopting a configuration
either interdigitating with the brush and standing end-on, or laying on top of the
brush, as in confetti scattered on the ground.  The additive will have to
determine which orientation (and which penetration depth) is mechanically
stable and the thermodynamic ground state orientation.
Figure~\ref{schematic} has a schematic indicating the two possibilities.

The additives I consider here are impenetrable to the polymer brush monomers and are relatively flat.
Additionally, I suppose a controllable interaction between the polymer brush and the solvent, and between the additive and the solvent.
In a real physical system, this additive could be a graphene sheet \cite{graphene}, suitably
decorated with side groups to control these interactions.
Additionally, fully exfoliated inorganic clay sheets with typical dimensions 
on the scale of the brush heights could be employed to create
a physically reasonable and technologically interesting instance of this
scenario \cite{clay_exfoliate}.
An even more interesting possibility is the ordering of carbon nanotubes (rather than
graphene sheets) perpendicular to the film normal 
creating  a smart surface that would
switch from a forest of trunks extending from the surface to a ``flattened forest'' in which all the nanotubes lay
along the surface.  This possibility will be explored in subsequent work,
as in Figure~\ref{schematic}(C).

I will thus estimate the free energy penalty to insert an isotropic plate of
typical dimension $L$ into a polymer brush in each of the two orientations.
This theory will make, necessarily, drastic assumptions simplifying the
situation.  The conclusions of this simple theory can 
be directly compared to a more calculationally 
intensive Scheutjens's and Fleer Self-Consistent-Field (SCF) 
lattice calculation \cite{sf_book}.
Below, I exhibit the generalization to 
the two-dimensional lattice model \cite{scf2d}
to encompass the presence of a freely-mobile, but interacting impenetrable
plate.
The most exciting result of this combined analytic and numerical approach
is the identification of a significant region in the polymer-plate
and polymer-solvent interaction space that results in the additives 
not only arranging themselves perpendicular to the brush, but also extending 
out past the {\em outer edge of the brush}.
This regions is accessible through changing the solvent quality or the
temperature, so there is the possibility that a ``smart'' surface, nominally
made of organic material, will present an extreme roughness consisting of
bare inorganic material plates, extending far above the surface, ready
to catch and hold several possible target materials.

\section{Simple Model}
Consider a square, rigid plate or side length $L$ that can interact first
with an end-grafted polymer brush consisting of molecules with a
degree of polymerization $N$, and a surface grafting density of $\sigma$ chains
per unit area.
A small molecule solvent is present, with Flory-Huggins interaction parameters
$\chi_p$ standing for the interaction between the {\em plate} and the solvent
and $\chi_b$ standing for the interaction of the {\em brush} and the solvent.
For this argument, I assume no energetic penalty or bonus for brush / plate contacts.
Certainly, when $\chi_b=\chi_p$ there is no tendency for the platelet 
additive to either adsorb at the brush or invade it significantly.
Increasing either $\chi_b$ or $\chi_p$ will trigger the absorption of the
additive.  On the one hand, the
plate in an increasingly bad solvent environment will be able to shield
its surface from bare contact with solvent by accumulating brush monomers
at its surface.  
Alternatively, as the solvent quality becomes harsher and harsher for the
brush, the plate will act as a compatibilizing agents, replacing some
brush-solvent contacts with brush-plate contacts.
The important question to address is how will the absorbed additive 
orient itself in the brush?

For well-soluble brushes ($0< \chi_b < \Theta =0.5$) and increasingly poor solvent conditions for the 
plate, the ultimate limit for the absorbed
plate will have the absorbed plate completely wet with brush monomers.
\begin{figure}
\centering
\includegraphics[scale=0.5]{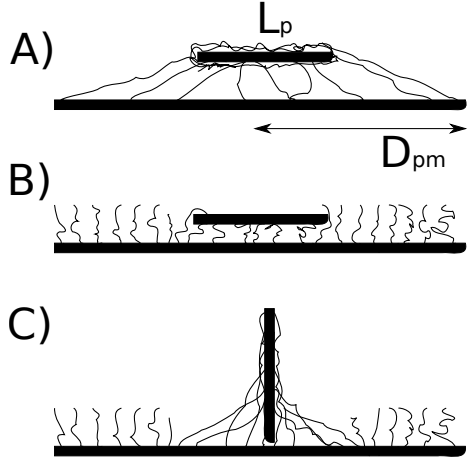}
\caption{
A) The plate draws monomers from grafted chains extending as far as $D_{pm}$, and fully covers the plate.
B) The solvent quality for the brush has worsened to the point that the plate dewets, exposing bare surface to solvent.
C) Intermediate solvent qualities, the plate orients perpendicular to the surface.}
\label{good_brush}
\end{figure}
The plate itself will then act as a concentration center for an 
induced micelle of brush polymers, similar in structure to an ``octopus''
or ``pinned micelle'' \cite{pinned,anna}.
As in Figure~\ref{good_brush}, the length $D_{pm}$ indicates the lateral
extent of the brush grafting surface contributing chains to the structure,
and is a function of brush grafting density, molecular weight, and
brush solvent quality.
Thus, we can estimate the free energy per plate for an initial ``confetti''
configuration in which both sides of the additive are wet with brush 
polymer:
\begin{equation}
F_{wet} = \sigma (L+D_{pm})^2   (L+ D_{pm})^2/ N.
\end{equation}
Here, the first term indicates number of chains involved in the structure,
and the second term indicates the stretching energy per chain in
the structure.  
As polymer-brush contacts are assumed to cost negligible free energy, 
and the plate is entirely sequestered from the solvent, there is no term
for the surface energy of the plate.
On the other hand, we can have a slightly lower stretching energy per 
chain, and fewer chains involved in the structure, in the perpendicular
orientation:
\begin{equation}
F_{perp} = \sigma (L + D_{pm}) L  ( L^2 +D_{pm}^2) / N,
\end{equation}
indicating that there will be a transition from the ``wetted confetti''
configuration to the ``perpendicular'' orientation when
\begin{equation}
L \approx D_{pm},
\label{three}
\end{equation}
apart from an uninteresting factor of order unity.
That is, when the plate itself is on the scale of the surface micelle its
presence induces, the transition from the wetted confetti to the perpendicular
orientation occurs.

The possibility exists that a further transition, from a perpendicular
orientation to a detwetted confetti configuration, can occur
when the plate has become so large that the enhanced stretching energy 
required for surface chains to extend all the way to the extreme
tip of the plate is balanced by the reduction in stretching energy of
a confetti orientation, with the outer surface of the plate wet with bad
solvent, and bare of polymer.
We can estimate the size of plate which causes such a transition.
In this case, the stretching energy of individual chains in the perpendicular
configuration are dominated by stretching over a length scale $L$:
\begin{equation}
F_{str} = L^2/ N
\end{equation}
and the surface energy penalty for the bare exposed confetti surface is (near the transition in Eq.~\ref{three})
\begin{equation}
F_{surf} = \chi_p L^2 ( L^2 \sigma)^{-1}.
\end{equation}
The transition, should it occur, would require these energies to be on the 
same scale:
\begin{equation}
L^2/ N = \chi_p L^2 ( L^2 \sigma)^{-1} \rightarrow L^\star = 
(N \chi_p / \sigma)^{1/2}.
\end{equation}
If $L< L^\star$, we can expect to increase $\chi_p$ without bound, and
yet never induce the transition to the bare confetti structure, whereas,
with $L > L^\star$, there is the possibility that the perpendicular 
orientation is preempted by the dewetting of the upper surface of the
confetti.
The broad possibilities for a good-solvent brush are thus enumerated.

When the brush is poorly soluble, I expect all values of $\chi_p$ will
result in the absorption of the plate.
Now, we have a relatively straightforward comparison of surface
energies in the three configurations.
The wetted confetti configuration has a free energy cost of
approximately
\begin{equation}
F_{wet} = (\sigma L a) L^2 / N + \chi_b L^2,
\end{equation}
where the first term counts up all the chains grafted at the outer
perimeter of the plate, and counts the cost for them to stretch across
the upper surface of the plate.
The second term counts the solvent-polymer surface energy.
The perpendicular orientation, on the other hand, does not require
stretching to coat both surfaces of the plate, and hence has a free
energy cost of
\begin{equation}
F_{perp} = L^2 \chi_b,
\end{equation}
as the brush-solvent contacts of the wetted configuration still occur for
the perpendicular configuration.
The expression for $F_{perp}$ is correct up to the point where $L$ exceeds the
layer thickness of the grafted layer: $h = \sigma N$.  When this occurs,
there is the possibility that a bare edge of the plate protrudes past the
edge of the brush, and we would then have to include the solvent-plate
energy cost:
\begin{equation}
F_{perp} = L^2 \chi_b + (L-h)L \chi_p, \mbox{ when } L > h.
\end{equation}
The detwetted confetti configuration will then have a free energy cost of
\begin{equation}
F_{confetti} = \chi_b L^2
\end{equation}
We thus expect that the wetted-confetti configuration to be stable from the
good-solvent regime up to the bad solvent regime, where its stretching
penalty destabilizes it to the perpendicular orientation.
Under these circumstances, as long as $L < h$, that transitions from 
perpendicular to dewetted confetti occurs when
\begin{equation}
\chi_b > \chi_p.
\end{equation}
Thus, increasing $\chi_b$ at fixed $\chi_p$ will eventually trigger a 
perpendicular-to-dewet transition. 
In this case, the flat-laying plates act as an anisotropic compatibilizer,
getting between the polymer and solvent.

In a situation where $L > h$, the balance of energies between the 
two configurations implies a transition when
\begin{equation}
\chi_b  L^2 + (L-h)L \chi_p = \chi_p L^2 
\end{equation}
or when
\begin{equation}
\chi_b L \approx h  \chi_p
\end{equation}
thus stabilizing the wetted configuration when the exposed perpendicular
segment of the plate has an area comparable to the whole plate area.

\section{Self-Consistent Field Results}
I have executed Scheutjens-Fleer \cite{sf_book} calculations in a system in which I assume translational invariance in one direction.
Thus, the plates I am formally considering are parallelepipeds with one small dimension (the lattice scale) one controllable dimension,
$L_p$, and one very long direction aligned in the $y$-axis.
The calculation lattice is $L_x= 60$ and $L_z=30$ sites in dimensions for the calculations I present.
In all that follows, I have kept the properties of the brush polymers fixed at $N=150$ monomers, and $\sigma=0.01$ grafted chains / surface lattice site.
Thus, a fully-packed polymer layer would have a size of $N \sigma= 1.5$ extent in the z-direction.
I consider additives with $L_p= 2, 4, 8$, and determine the equilibrium orientation of the additive by ``holding'' it a distance
$z_p$ above the grafting surface in each orientation, and determining which $z_p$ and corresponding orientation minimizes the overall
system free energy for a single additive.

\begin{figure}[t!]
\centering
\includegraphics[scale=0.8]{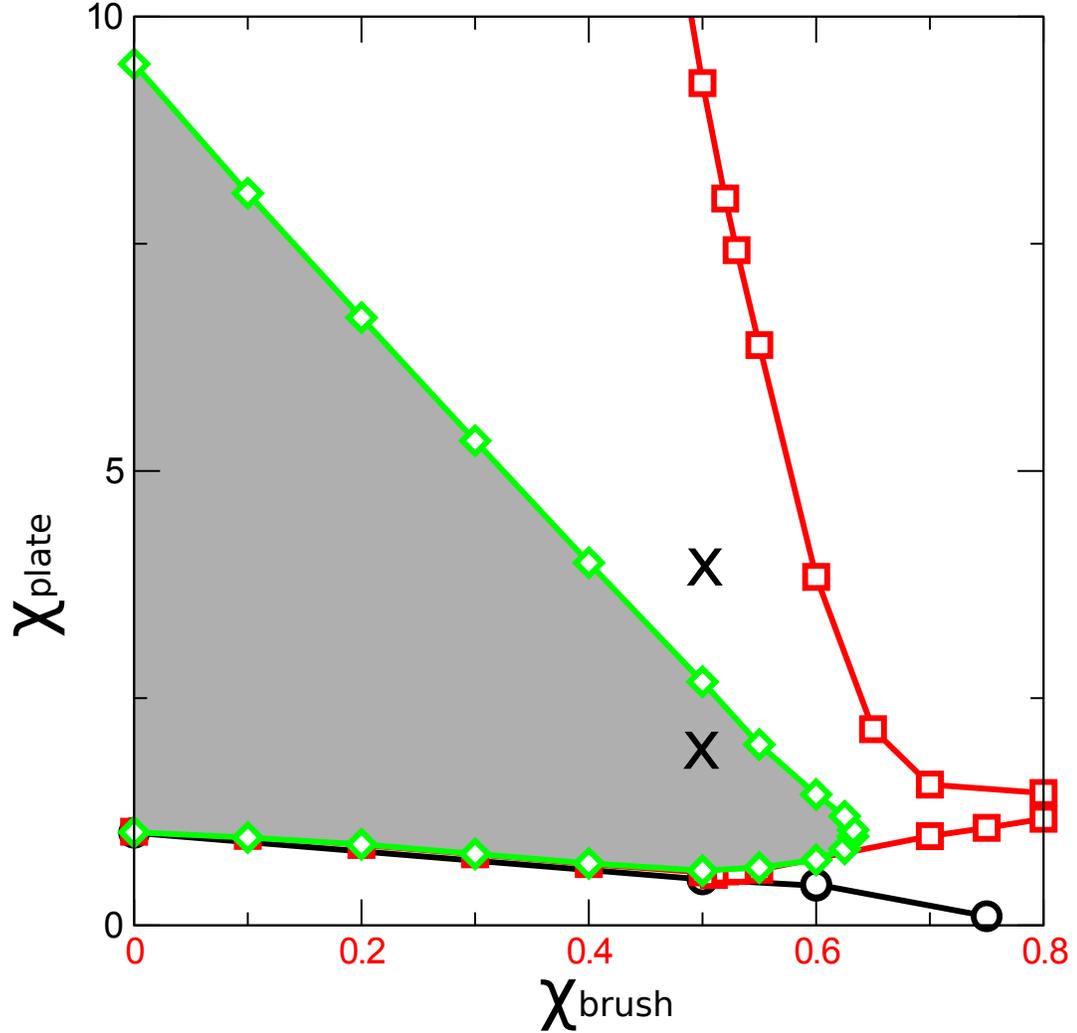}
\caption{ 
When the plate is on the order of the fully compressed layer $L_z=2$, black circles, the additives are ``parallel'' below the line, and ``perpendicular''
above it.
For $L_p=4$, there is a large region with perpendicular orientation (to the left and above the red square line).
When the additive is much longer than the fully compressed polymer layer, $L_p= 8$, compared to a layer size of $\approx 2$, the region of perpendicular orientations is shaded on the diagram.  Here, ``reentrant'' transitions from parallel-perpendicular-parallel occur generically.
The two ``X'''s mark parameters corresponding to the brush volume fraction profiles in Fig.~\ref{both}.
}
\label{scf}
\end{figure}
The additive is modeled as a region of $L$ lattice sites completely filled with a volume fraction of $\phi_p=1$.
There are two Flory-Huggins parameters in the problem.  One, $\chi_b$ is the interaction
parametner between the brush segments and the monomeric solvent.
The other, $\chi_p$ is the plate-solvent interaction parameter.  I have assumed that the interaction between polymer and plate
has $\chi=0$, as in the above.
Thus, the solvent is the effective medium of interaction between the brush and the plate, modified by the connectivity of the 
polymer chains and the orientation of the additive.

\begin{figure}
\centering
\includegraphics[scale=0.8]{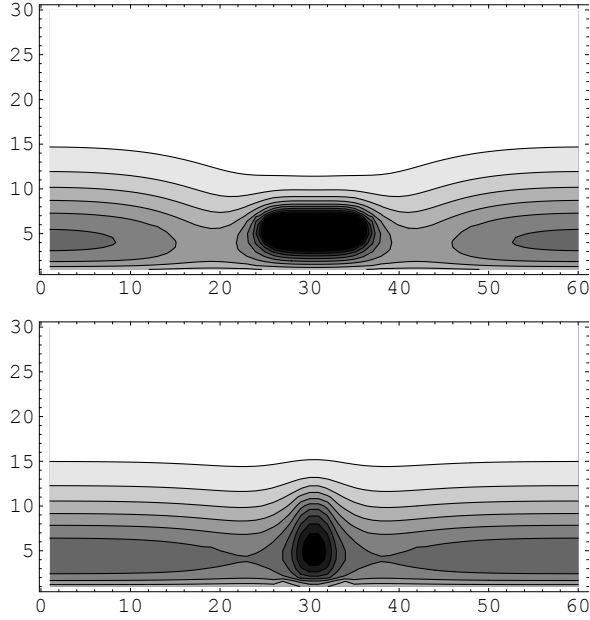}
\caption{ 
The upper panel corresponds to $\chi_b=0.5$ and $\chi_p=4.0$ (the upper ``X'' in Fig.~\ref{scf}.
The lower panel corresponds to $\chi_b=0.5$ and $\chi_p=2.0$ (the lower ``X'' in Fig.~\ref{scf}.
}
\label{both}
\end{figure}
Figure~\ref{scf} shows the results of the calculation, showing the region where the perpendicular orientations exist.
When $L_p=2$, there is a small region of parallel orientations for both small $\chi_b$ and $\chi_p$, but most of the parameter space is dominated by the 
perpendicular orientation.
For plates a bit longer, $L=4$, the scenario in the above section plays, out.  
There are regions in which transitions from parallel-perpendicular-parallel occur (for increasing $\chi_b$ at $\chi_p=0.6$, but when $\chi_b<0.4$ even the
unrealistically high value of $\chi_p=10$ is insufficient to cause the additive to lay down.
For much longer additives, $L_p=8$, no matter what the plate-solvent interaction, there is always a point at which the polymer layer ejects the 
perpendicular plate, which then 
lays on top of the brush.  In this scenario, the plate is wet with polymer.

\section{Discussion}
Several means of controlling surface properties of a polymer brush seem possible through this mechanism, and the collective orientation behavior for
particles in the brush is worth of further study. 
At a minimum, {\it in situ} control of the brush layer as a surface in a many-step, hierarchical nanoscopic self-organization for device applications seems 
possible.  
With a change in solvent quality, the lateral area fraction of the brush covered by the additives can change discontinuously in going from the perpendicular
to the dewetted parallel configuration.
When many perpendicularly oriented additives are present, it is possible for the additives to behave as a two-dimensional confined ``clay'' material, with the
possibility of liquid crystal phases in their lateral orientation, and perhaps a transition from a dilute ``ex-foliated'' structure to more densely packed 
configurations reminiscent of ceramic clays (albeit nanoscopically confined, end-on clays).

Also, I have made no attempt here to describe the dynamics of the additive when in the perpendicular configuration.
It is possible that this configuration will change the effective dynamic boundary condition at the polymer surface, so that a shearing field for the solvent
could have additive-controlled (and therefore switchable) transitions from stick to slip dynamic boundary conditions.

Even more interesting to me are the possibilities opened when the additives have a more complex geometry.  
Single-walled carbon nanotubes and nanowires are examples.  
A very interesting set of complex materials with exotic geometries can be constructed by designing DNA strands to form T-bars, tetrahedra, other complex
shapes \refto{dna}.
Such objects have the capability of not only orienting in multiple ways in the polymer layer, 
but of changing their conformations in response
to the stress the grafted polymers put upon the additives.

\section{Conclusion}
Flat, hard additives added to an end-grafted polymer brush can orient either perpendicular or parallel to the polymer layer.
The orientation is controlled by the ratio of the layer thickness and the size of the additive and the solvent interactions between the solvent and the brush
and the solvent and the additive.
An interesting perpendicular orientation in which bare additive is exposed far from the surface occurs in intermediate solvent conditions.

\end{document}